# CRITERIA OF TURBULENT TRANSITION IN PARALLEL FLOWS

HUA-SHU DOU

*Temasek Laboratories, National University of Singapore,
Singapore 117508*
*tsldh@nus.edu.sg; huashudou@yahoo.com*

BOO CHEONG KHOO

*Department of Mechanical Engineering, National University of Singapore,
Singapore 119260*
*mpekbc@nus.edu.sg*

**Abstract**   Based on the energy gradient method, criteria for turbulent transition are proposed for pressure driven flow and shear driven flow, respectively. For pressure driven flow, the necessary and sufficient condition for turbulent transition is the presence of the velocity inflection point in the averaged flow. For shear driven flow, the necessary and sufficient condition for turbulent transition is the existence of zero velocity gradient in the averaged flow profile. It is shown that turbulent transition can be effected via a singularity of the energy gradient function which may be associated with the chaotic attractor in dynamic system. The role of disturbance in the transition is also clarified in causing the energy gradient function to approach the singularity. Finally, it is interesting that turbulence can be controlled by modulating the distribution of the energy gradient function.

***Keywords*:** Turbulent Transition, Criteria, Energy gradient, Role of disturbance, Singularity.

# 1. Introduction

Turbulence is one of the most difficult problems in classical physics and mechanics. The origin of turbulence is an intensive research topic for many years. It has been observed in experiments that turbulent transition depends on the Reynolds number and the amplitude of disturbance. The disturbance amplitude is found to scale with Re by an exponent of -1 with transversal injection disturbance [1]. This phenomenon has been successfully modeled by the energy gradient method [2-5]. There is good agreement between the energy gradient method and experiments for parallel flows in three aspects: the scaling of the disturbance threshold with Re, the location of the maximum disturbance in the transition and the critical value of the energy gradient function. It also explained the possible mechanism of receptivity to free-stream disturbance and the mechanism of self-sustenance of turbulence in boundary layer flows which is absent in pipe and channel flows. In this study, based on the energy gradient method [2-5], criteria for turbulent transition in parallel flow are proposed for both pressure driven and shear driven flows.

# 2. Energy Gradient Theory: Re-visited

Dou [2] proposed an energy gradient method for the purpose of clarifying the mechanism of flow instability and turbulent transition. It is demonstrated that the relative magnitude of the total mechanical energy of fluid particles gained and the energy loss along streamline due to viscous friction in a disturbance cycle determines the disturbance amplification or decay. For a given base flow, a stability criterion is expressed as [2-5],

$$K \frac{v'_m}{U} < Const, \quad K = \frac{(\partial E / \partial n)}{(\partial H / \partial s)}, \tag{1}$$

where $K$ is a dimensionless field variable (function) and expresses the ratio of the gradient of the total mechanical energy in the transverse direction and the loss of the total mechanical energy in the streamwise direction. Here, $E = p + 0.5\rho V^2$ is the total mechanical energy per *unit volumetric fluid*, $s$ is along the streamwise direction, $n$ is along the transverse direction, $H$ is the energy loss per *unit volumetric fluid* along the streamline, $\rho$ is the fluid density, u is the streamwise velocity of main flow, U is the characteristic velocity, and $v'_m$ is the amplitude of disturbance velocity. Since the magnitude of *K* is proportional to the global Reynolds number ($\text{Re} = \rho U L / \mu$) for a given geometry [2], the criterion of Eq.(1) can be written as [3-5],

$$\text{Re} \frac{v'_m}{U} < Const \quad \text{or} \quad (\frac{v'_m}{U})_c \sim (\text{Re})^{-1}. \tag{2}$$

This scaling has been confirmed by careful experiments observed for pipe flow and boundary layer flow, and this result is in agreement with the asymptotic analysis of the Navier-Stokes equations (for $\text{Re} \to \infty$) [3-5].

In Eq.(1), K corresponds to the *base flow* where the flow parameters are calculated for the smooth laminar flow at given Re. When a disturbance with finite amplitude is imposed on the flow, the distribution of K of the *averaged flow* is different from that of the smooth laminar flow without disturbance imposed.

## 3. Criteria for Turbulent Transition

As is well known, a laminar flow is smooth in which the motion of fluid can be described by deterministic equations whereas a turbulent flow is of disorder and chaotic. The question is how a smooth laminar flow becomes turbulent/chaotic flow in which the motion of fluid becomes largely non-deterministic? What is the role of disturbance in the transition? From Eq.(1), it is observed that the value of K at transition tends to infinity when the disturbance is infinitesimal. In other words, a smooth laminar flow can transit to turbulence via singularity in the K function. This is reasonable from the viewpoint of mathematics because the behavior of singularity can be indefinite. If the disturbance is increased to a non-zero value, the path towards singularity for the transition is still needed to be consistent with the indefinite behaviour of turbulent motion. In other words, it is necessary that the energy gradient function of *averaged flow* with imposed disturbance must be singular for turbulent transition occurrence, i.e., $K = \infty$.

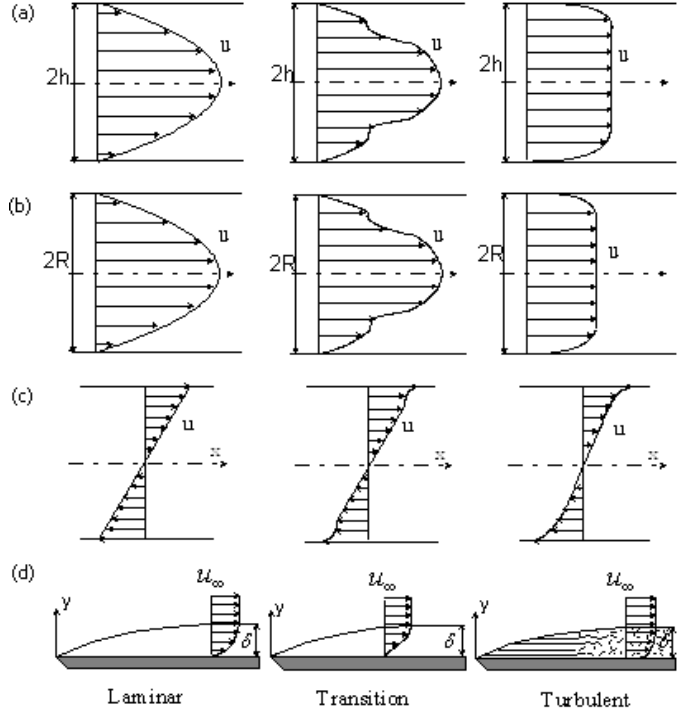

Fig.1 Turbulent transitions in parallel shear flows. (a) Plane Poiseuille flow; (b) Pipe Poiseuille flow; (c) Plane Couette flow; (d) Boundary layer flow on a flat plate.

Since $\partial E/\partial n$ in Eq.(1) is always limited, we have $\partial H/\partial s = 0$ for K to be singular. For pressure driven flows, from Navier-Stokes equation for parallel flows [2], we obtain the Eq.(3) for two-dimensional or axisymmetric flows

$$\frac{\partial H}{\partial s} = \mu\left(\frac{\partial^2 u}{\partial y^2}\right) = 0 \quad \text{or} \quad \frac{\partial H}{\partial s} = \mu\frac{1}{r}\frac{\partial}{\partial r}\left(r\frac{\partial u}{\partial r}\right) = 0. \tag{3}$$

Therefore, for pressure driven flow, the necessary condition for turbulent transition is the presence of velocity inflection of the *averaged flow profile*. For shear driven flows, $\partial H/\partial s = 0$ can be obtained from analysis of energy process [4], i.e.,

$$\frac{\partial H}{\partial s} = \frac{\tau}{u}\frac{\partial u}{\partial y} = \frac{1}{u}\mu\left(\frac{\partial u}{\partial y}\right)^2 = 0. \tag{4}$$

Therefore, for shear driven flow, the necessary condition for turbulent transition is the existence of zero velocity gradient on the velocity profile of the *averaged flow*.

The proof for sufficiency of these criteria is obvious from Eq.(1) since an infinite value of K of averaged flow will significantly amplify any disturbance present in the flow.

## 4. Comparison with Experiments

The validity of these criteria can be compared with available experiments (Fig.1). For plane Poiseuille flow, the experiments by Nishioka et al showed that the averaged flow displays an inflection when the disturbance is large which is necessary as the transition is approached [6]. For pipe Poiseuille flow, the experiments by Nishi et al [7] showed that the averaged flow indicates an inflection when the transition occurs. Wedin and Kerswell's simulation also displays an inflection under a disturbance of travelling waves on the averaged velocity profile [8]. For plane Couette flow, there is still no experiment on the velocity profile during the transition. However, the simulation by Lessen and Cheifetz showed the appearance of the zero velocity gradient near the moving wall [9]. For boundary layer flow on a flat plate, it is found in experiments [10, 11] that turbulence is originated on the wall, and it is known that there is velocity inflection at the wall.

## 5. Conclusions

Criteria for turbulent transition in parallel flows are proposed for pressure driven flows and shear driven flows, respectively. These criteria have been compared with the available experiments and the validity of these criteria is confirmed. It is shown that laminar flow transits to turbulence via a singularity of energy gradient function which is believed to be associated with the chaotic attractor in dynamic system. The role of disturbance is to promote or shift the averaged flow towards this singularity.